\documentclass[twocolumn,showpacs,preprintnumbers,epsfig,amsmath,amssymb]{revtex4}
 
\usepackage{graphicx}
\usepackage{dcolumn}
\usepackage{bm}
 
\begin{document}
 
\preprint{}
\title{Specific heat anomaly in the d-density wave state and emergence of incommensurate 
orbital antiferromagnetic order}
\author{Hae-Young Kee and Yong Baek Kim}
\affiliation{Department of Physics, University of Toronto, Toronto, Ontario, 
Canada M5S 1A7}
\date{\today}

\begin{abstract}
We study the effect of finite chemical potential on the d-density wave 
state that has been proposed to explain the pseudogap phenomena in 
underdoped cuprates.
We find that the specific heat anomaly at the transition temperature,
below which the d-density wave state forms, gets weaker when finite
chemical potential is introduced. 
This provides a useful ground for the proper interpretation of 
the specific heat measurement in regard to the existence of
the d-density wave state below the pseudogap temperature.  Further 
increase of the chemical potential 
leads to an incommensurate orbital antiferromagnetic state 
before the system eventually turns into the normal state. 
This is an inhomogeneous state
characterized by novel charge ordering and an analog of  
the Fulde-Ferrel-Larkin-Ovchinnikov (FFLO) state in superconductivity.
\end{abstract}
 
\maketitle

\paragraph{Introduction}
One of the recent interesting proposals made in high 
temperature superconducting cuprates is that the so-called 
d-density wave state is responsible for the pseudogap 
phenomena.\cite{sudip00,hykee}
The d-density wave state is a particle-hole paired state
with a finite momentum ${\bf Q}_0=(\pi,\pi)$.
The particle-hole pair wavefunction has a d-wave symmetry, 
$f({\bf k}) \propto \cos{k_x}-\cos{k_y}$, that occurs due to the 
formation of staggered currents circulating in the
square lattice. Therefore, the d-density wave state breaks the 
translational and time reversal symmetries.\cite{marston88,nayak}

Since the d-density wave state is an orbital antiferromagnetic 
state, there should be an elastic neutron scattering signal 
at the momentum ${\bf Q}_0$. This is a direct consequence
of the broken symmetry. It was argued that the recent neutron 
scattering experiment is consistent with the existence 
of the d-density wave state in underdoped cuprates \cite{mook01,sudip01}.
On the other hand, there is no clear experimental evidence of the 
broken symmetry state in thermodynamic quantities, such as
specific heat at $T^*$ where the pseudogap develops.
It has been regarded that this is at odds with the proposal, because 
one may expect that there should be a clear specific heat
jump due to the translational symmetry breaking.
In fact, in the case of the half-filled band, the normalized specific 
heat jump for the d-density wave state is the same as that of the 
d-wave superconducting state within the mean field theory.
This is simply because these states have the same form of the
quasiparticle spectrum, 
$E_{\bf k}=\sqrt{\epsilon_{\bf k}^2+\Delta_{\bf k}^2 }$ 
when the chemical potential is zero ($\mu=0$).
Here $\Delta_{\bf k}=\Delta_0 \cos{(2\theta_{\bf k})}$
is the d-wave gap in the d-density wave state or in the 
d-wave superconductor.
The specific heat measurements \cite{loram} 
show a clear evidence of the phase transition at $T_c$,
while there is no anomaly reported at $T^*$.

This conclusion should be, however, reexamined
when the system is 
away from half-filling. In the presence of the finite chemical 
potential, $\mu$, the quasiparticle 
spectrum is quite different for two cases.
In the d-wave superconductor, the quasiparticle dispersion is given 
by $E_{\bf k}=\sqrt{(\epsilon_{\bf k}-\mu)^2+ \Delta_{\bf k}^2 } $,
while in the d-density wave state it is given by
 $E_{\bf k}=-\mu \pm \sqrt{\epsilon_{\bf k}^2+ \Delta_{\bf k}^2 } $.
The difference comes from the fact that one is a particle-particle 
paired state and the other is a particle-hole paired state.

In this paper, motivated by the above observations, 
we study the effect of the finite chemical potential
on the commensurate orbital antiferromagnetic state dubbed 
``d-density wave state''. 
Using a BCS-like Hamiltonian for the d-density
wave state, we show that 
{\bf 1)} the specific heat anomaly at the transition temperature,
$T=T_d$, due to the d-density wave ordering
gets weaker as the chemical potential increases.
Therefore, when the chemical potential is appreciably large, 
it will be hard to determine the existence
of the d-density wave state by measuring the specific heat.
In view of the importance of the finite chemical potential, 
we ask whether there is a phase transition driven by the 
chemical potential. We find that 
{\bf 2)} the stability of the d-density wave state gets weaker
as the chemical potential increases, and beyond a critical
value, an incommensurate orbital antiferromagnetic state occurs
in a range of chemical potential. 
It is an inhomogeneous state 
characterized by charge ordering in contrast to the
d-density wave state with uniform charge density and
an analog of the FFLO state in superconductivity.\cite{iddw}
Certain choices of the incommensuration correspond to
the stripe patterns of the charge ordering similar to
those discussed in the context of the spin density wave
in the cuprates.\cite{kivelson}

\paragraph{Role of the chemical potential}
The d-density wave state is a particle-hole paired state 
where the band dispersion of the particles and holes satisfy 
$\epsilon_{\bf k}= -\epsilon_{{\bf k}+{\bf Q}_0}$ with
${\bf Q}_0=(\pi,\pi)$.
Within mean field approximation, the d-density wave state can 
be stable in an extended Hubbard model at half-filling, i.e., 
$\mu=0$, where the particle-hole symmetry exists.\cite{nayak}

When the chemical potential becomes finite, $\mu \neq 0$,
the number of particles is not equal to that of holes.
Therefore, the chemical potential suppresses the tendency 
of paring and effectively acts as a pair breaker.
In other words, the nesting condition gets weaker with finite 
chemical potential, so does the stability of the d-density 
wave state.
In fact, the chemical potential can be regarded as an analog 
of the Zeeman magnetic field in superconducting states,
where the Zeeman field splits the up- and down-spin bands and
acts as a pair breaker.
We will show below that the gap equation of the d-density wave
state with finite $\mu$, Eq.\ref{gap}, is the same as that of
the d-wave superconducting state with a finite Zeeman 
field.\cite{maki96,yang98}
Once we realize this simple analogy, it is clear that the 
specific heat anomaly 
will be reduced by introducing the chemical potential.
This is because the Fermi surface is no longer a point, but 
there will be a hole pocket whose size depends on the 
magnitude of $\mu$. There were similar studies in the spin density 
wave showing that the specific heat jump gets smaller 
when the nesting condition gets weaker.\cite{huang92}

\paragraph{Specific heat anomaly in the d-density wave state}
In order to compute the specific heat jump at the transition temperature, 
$T_d$, we use the following BCS-like Hamiltonian.
\begin{eqnarray}
H &- &\mu N =
\sum_{{\bf k},\sigma}(\epsilon_{\bf k}-\mu) c^{\dagger}_{{\bf k},\sigma}
c_{{\bf k},\sigma}   
\nonumber\\
&+ & \sum_{{\bf k},{\bf k}^{\prime},{\bf Q},\sigma,\sigma^{\prime}} 
g_{{\bf k}{\bf k}^{\prime}{\bf Q}} \ 
c^{\dagger}_{{\bf k},\sigma}
c_{{\bf k}+{\bf Q},\sigma}
c_{{\bf k}^{\prime}-{\bf Q},\sigma^{\prime}}
c^{\dagger}_{{\bf k}^{\prime},\sigma^{\prime}},
\end{eqnarray}
where $\epsilon_{\bf k}$ is the band dispersion and satisfies
$\epsilon_{\bf k}=-\epsilon_{{\bf k}+{\bf Q}_0}$.
$g_{{\bf k}{\bf k}^{\prime}{\bf Q}}$ is the pairing
potential that takes the following form when ${\bf Q}={\bf Q}_0$.
\begin{equation}
g_{{\bf k}{\bf k}^{\prime}{\bf Q}_0}= -g \cos{(2\theta_{\bf k})}
\cos{(2\theta_{{\bf k}^{\prime}})}.
\end{equation}

The gap equation is obtained in the mean field approximation as follows.
\begin{eqnarray}
1&=&\frac{g N(0)}{2} \int_0^{\omega_c} d{\epsilon} \int_0^{2\pi}
\frac{ d\theta}{2\pi}
\frac{\cos^2{2\theta}}
{\sqrt{\epsilon^2+\Delta^2 \cos^2{2\theta}}}
\nonumber\\
&\times &
\left[ \tanh\left( \frac{E_{\bf k}-\mu}{2 T} \right)
+ \tanh\left( \frac{E_{\bf k}+\mu}{2 T} \right) \right],
\label{gap}
\end{eqnarray}
where $E_{\bf k}^2=\epsilon_{\bf k}^2+\Delta^2 \cos^2{2\theta}$
and we define $\mu$ to be positive.
Here $N(0)$ is the density of state at $\epsilon=0$,
and we have assumed that there is no singularity in $N(0)$.\cite{note}

Using the standard procedure, one can obtain the following formula
for the transition temperature $T_d(\mu)$ at finite chemical potential.
\begin{equation}
\ln{ \frac{T_{d0}}{T_d}} =
{\rm Re} \psi \left(\frac{1}{2}+\frac{i \mu}{2 \pi T_d} \right)
-\psi\left(\frac{1}{2} \right),
\end{equation}
where $T_{d0}$ is the transition temperature when $\mu=0$
and $\psi(x)$ is the di-gamma function.
As expected, $T_d(\mu)$ decreases as the chemical potential 
increases. 

Now let us evaluate the specific heat jump at $T=T_d$.
Since we are interested in the specific heat anomaly at $T=T_d$,
we expand the gap equation in $\Delta(\mu,T)/\pi T$.
Using the conventional method described in \cite{abrikosov}, 
the specific heat jump at $T_d$ can be computed as 
\begin{eqnarray}
C_{ddw}(T_d) &-& C_N(T_d)
= N(0)  \frac{16 \pi^2 T_c}
{ 3 {\rm Re} \zeta{(3,1/2+i \mu/2 \pi T_d)} }
\nonumber\\
&\times&
\left[ 1+\frac{ \mu}{2 \pi T_d} {\rm Im}
\zeta{(2,\frac{1}{2}+\frac{i \mu}{2 \pi T_d})} \right]^2,
\end{eqnarray}
where $\zeta{(s,a)}=\sum_{n=0}^{\infty} 1/(n+a)^s$.
Here $C_{ddw}$ and $C_N$ represent specific heat of the
d-density wave state and normal state respectively.  

The specific heat coefficient $\gamma_{ddw}=C_{ddw}/T$
at $T=T_d$ is shown in Fig. 1 as a function of $\mu/T_d$, 
where $\gamma_N$ in the normal state is set to be unity ($\gamma_N=1$).
As shown in Fig. 1, the specific heat anomaly, 
$\delta \gamma_{ddw}/\gamma_N
=(\gamma_{ddw}-\gamma_N)/\gamma_N$ gets smaller
as $\mu$ increases.
Now let us compare the absolute value of the specific heat anomaly,
$\delta \gamma_{ddw}/\gamma_N$, with that of the d-wave superconducting 
state, $\delta \gamma_{sc}/\gamma_N = 0.9$. 
In the case of the d-density wave state, when 
$\mu= 0.6 T_d = 0.56 T_{d0}$, $\delta \gamma_{ddw}/\gamma_N = 0.26$.

\begin{figure}[h]
\vspace{-2.5cm}
\includegraphics[height=8cm,width=9cm,angle=0]{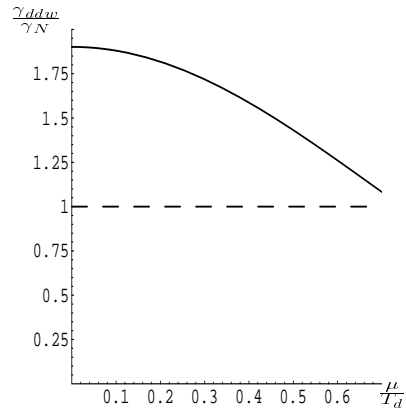}
\caption{The specific heat coefficient, $\gamma_{ddw}/\gamma_N$ is
shown as a function of $\mu/T_d$. The dashed line represents 
$\gamma_N$ in the normal state, set to be unity.
Notice that $\mu/T_d$ is a monotonically increasing
function of $\mu/T_{d0}$.
\label{figure:1}}
\end{figure}
Thus, the specific heat jump associated with the d-density wave
state can be quite small compared with that of the d-wave superconducting 
state when the chemical potential is finite.
Notice that our results are obtained in the mean field theory and 
the anomaly can be even smaller once we take into account the 
fluctuation effects.  

\paragraph{Phase transition}
Provided that the chemical potential in
the d-density wave state plays the same role as the Zeeman magnetic
field in superconductors, let us first review the phase transition 
driven by a Zeeman field in superconductors.
It is clear that sufficiently strong Zeeman field will 
eventually drive a phase transition into the normal state. 
The nature of the transition is determined by the competition
between two energy scales; the condensation energy 
$\frac{1}{2} N(0) \Delta^2$ and the magnetic energy
$\chi H^2$ where $\chi$ is the spin susceptibility proportional 
to the density of state $N(0)$.
This competition leads to the phase transition between the BCS state 
and normal state at a critical Zeeman field 
$H_c = \Delta/(\sqrt{2} \mu_B)$ ($\mu_B$ is the Bohr magneton) 
for s-wave superconductors.
On the other hand, it has been also known that, for a range of 
intermediate strength of the Zeeman field around $H_c$, it is more 
favorable to pair up- and down-spin electrons across the spin-split 
Fermi surfaces and form an inhomogeneous superconducting state.
This is called the Fulde-Ferrel-Larkin-Ovchinnikov (FFLO) 
state.\cite{larkin,fulde}
The phase transition from the BCS superconducting state to
the FFLO state is the first order transition for both
s-wave and d-wave superconductors\cite{larkin,maki96,yang98}.

The d-density wave state with finite chemical potential should 
experience the same kind of phase transition due to the existence
of two competing energy scales.
The condensation energy favors 
the particle-hole paired state. As the chemical potential increases,
however, the normal state is favored by $N(0) \mu^2$.
Therefore, the qualitative phase diagram should be similar to those 
obtained in superconductivity\cite{yang98,shimahara94}.
What is really interesting is that there is a window of chemical
potential where a novel phase, an analog of the FFLO state, is favorable 
over the commensurate orbital antiferromagnetic
and normal states. This phase is the incommensurate orbital 
antiferromagnetic state where the particle-hole pairing occurs with 
a momentum ${\bf Q} = {\bf Q}_0+{\bf q}$, where 
$|{\bf q}| \propto \mu$ at $T=0$.  
Similar studies were undertaken in the context of the excitonic 
insulator and an inhomogeneous phase with the s-wave symmetry 
was identified. \cite{gorkov00}

\paragraph{Incommensurate orbital antiferromagnetic state}
The incommensurate orbital antiferromagnetic order
arises when a staggered current is circulating in the plane 
but its periodicity is incommensurate 
with the underlying square lattice.

\begin{figure}[h]
\includegraphics[height=6cm,width=8cm,angle=0]{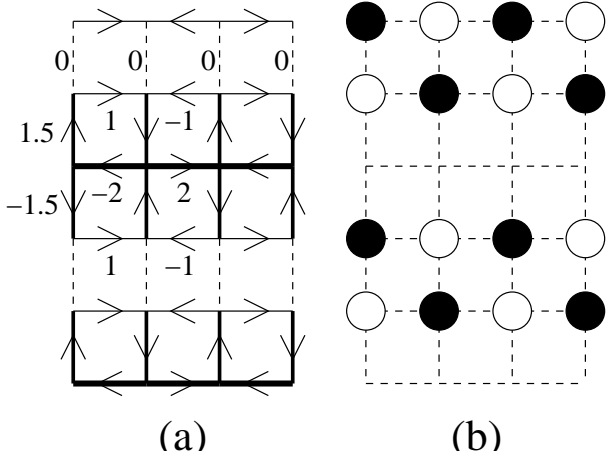}
\caption{\label{figure:2} (a) An example of the incommensurate
orbital antiferromagnetic state with ${\bf q}=(0,\pi/3)$.
The numbers represent the relative strength of each bond current.
Since $q_x=0$,  the current is alternating
with the same strength along the x-direction.
(b) The charge modulation in real space for ${\bf q}=(0,\pi/3)$.
The black and white circles represent the amplitudes of the charge
modulation proportional to $\sin{(\pi/3)}$ and $-\sin{(\pi/3)}$
respectively. There is no charge modulation at the empty sites.}
\end{figure}

{\bf Current modulation:}
It is straightforward to obtain the expectation value of the 
bond current in the incommensurate orbital antiferromagnetic state,
$\langle J_{{\bf r} {\bf r}^{\prime}} \rangle 
= i \langle c^{\dagger}_{\bf r} c_{{\bf r}^{\prime}}
- c^{\dagger}_{{\bf r}^{\prime}} c_{{\bf r}} \rangle $.
Taking ${\bf r}^{\prime}={\bf r}+{\hat x}$
and  ${\bf r}^{\prime}={\bf r}+{\hat y}$,
the current modulation is obtained as 
\begin{eqnarray}
& & \hspace{-1.0cm} \langle J_{{\bf r},{\bf r}+ {\hat x}}
 - J_{{\bf r},{\bf r}+ {\hat y}} \rangle
=  i  \sum_{\bf k} 
 \langle c^{\dagger}_{\bf k} c_{{\bf k}+{\bf Q}_0+{\bf q}}  \rangle
\nonumber\\
& \times & 
\{ \ [f({\bf k},{\bf q})-i g({\bf k},{\bf q})] 
\cos{[({\bf Q}_0+{\bf q}) \cdot {\bf r}]}
\nonumber\\
& & - [g({\bf k},{\bf q})+i f({\bf k},{\bf q})] 
\sin{[({\bf Q}_0+{\bf q}) \cdot {\bf r}]} 
+c.c. \}.
\end{eqnarray}
Here $f({\bf k},{\bf q})$ and $g({\bf k},{\bf q})$ are given by  
\begin{eqnarray}
f({\bf k},{\bf q}) & = & \cos{k_x} - \cos{k_y} \cr
& & + \cos{(k_x+q_x)} - \cos{(k_y+q_y)},
\nonumber\\
g({\bf k},{\bf q}) & = & \sin{k_y} -\sin{k_x}
\nonumber\\
& & + \sin{(k_x +q_x)} -\sin{(k_y +q_y)}.
\label{internal}
\end{eqnarray}

The following order parameter of a particle-hole paired state with the
momentum ${\bf Q}={\bf Q}_0+{\bf q}$ is the simplest choice for
the incommensurate orbital antiferromagnetic order. 
\begin{equation}
 \langle c^{\dagger}_{\bf k} c_{{\bf k}+{\bf Q}_0+{\bf q}} \rangle 
= i \frac{1}{2} \Delta_{Q_0+q} f({\bf k},{\bf q}),
\label{order}
\end{equation}
where
$f({\bf k},{\bf q})$ is given by  Eq. (\ref{internal}).
This order parameter of the particle-hole pairing is an analog of 
the FFLO order parameter of the particle-particle pairing.
It is also reduced to the order parameter of the d-density wave 
state when ${\bf Q}={\bf Q}_0$. 

The order parameter, Eq. \ref{order}, leads to
the following current pattern in real space.
\begin{eqnarray}
\langle J_{{\bf r},{\bf r}^{\prime}} \rangle
&=& (-1)^{\bf r} \Delta_{Q_0+q}
\nonumber\\
&\times& \left[ \delta_{x^{\prime},x+1}\delta_{y^{\prime},y}
\{  \cos{({\bf q}\cdot {\bf r})}
+ \cos{({\bf q}\cdot {\bf r}+q_x)} \} \right.
\nonumber\\
&+& \delta_{x^{\prime},x-1}\delta_{y^{\prime},y}
\{  \cos{({\bf q}\cdot {\bf r})}
+ \cos{({\bf q}\cdot {\bf r}-q_x )} \} 
\nonumber\\
&-&\delta_{y^{\prime},y+1}\delta_{x^{\prime},x}
\{  \cos{({\bf q}\cdot {\bf r})}
+ \cos{({\bf q}\cdot {\bf r}+q_y )} \} 
\nonumber\\
&-& \left. \delta_{y^{\prime},y-1}\delta_{x^{\prime},x}
\{  \cos{({\bf q}\cdot {\bf r})}
+ \cos{({\bf q}\cdot {\bf r}-q_y )} \}
\right]
\end{eqnarray}
One example of the incommensurate staggered
current patterns is shown in Fig. 2 (a), where 
${\bf q}=(0,\pi/3)$. Different choices of the ordering
vector ${\bf Q}$ lead to different patterns of 
the incommensurate staggered currents.
For ${\bf q}=(0,\pi/3)$, the current density has the modulation 
with the period $[2a, 3a]$ in the $x$- and $y$-directions, where 
$a$ is the lattice constant.
Notice that there exists a set of the vanishing bond currents, 
$J_{{\bf r},{\bf r}+{\hat y}}=0$, that 
are aligned along the $x$-direction. This ``bond-centered'' stripe-like
structure leads to the charge modulation (computed in the next section)
that has the same periodicity as that of the current, as shown in Fig. 2 (b). 

On the other hand, when ${\bf q}=(0,\pi/4)$, a set of the vanishing bond 
currents in the $x$-direction, $J_{{\bf r},{\bf r}+{\hat x}}=0$, exists.
This corresponds to the ``site-centered''
stripe-like structure along the $x$-direction in the current pattern.
As a result, the periodicity of the current is $[2a,8a]$, while that 
of the charge density is $[2a,4a]$. That is, the periodicity of the current 
along the $y$-direction is twice larger than that of the charge.

{\bf Charge ordering in the incommensurate orbital antiferromagnetic state:}
The staggered current pattern in the d-density wave state gives 
{\it no} charge modulation due to
$\sum_{{\bf k} \in RBZ} (\cos{k_x}-\cos{k_y}) = 0$,
where $RBZ$ stands for the reduced Brillioun zone. 
This property distinguishes the d-density wave state from the 
ordinary charge density wave state.

In the incommensurate orbital antiferromagnetic state,
we find that there {\it exists} a charge modulation.
As an example, let us compute the charge density when 
the staggered current pattern is given by Fig. 2 (a),
where ${\bf q}=(0,\pi/3)$.
The local charge density can be obtained from
\begin{eqnarray}
&& \hspace{-1.5cm} \langle \psi^{\dagger}({\bf r}) \psi({\bf r)} \rangle
= \sum_{\bf k} \sum_{{\bf k}^{\prime}}
\langle c^{\dagger}_{\bf k} c_{{\bf k}^{\prime}} \rangle
\exp{[i({\bf k}-{\bf k}^{\prime}) \cdot {\bf r}]}
\nonumber\\
&=& n_0 + \Delta_{Q_0+q}
h(q_y) \sin{[({\bf Q}_0+{\bf q}) \cdot {\bf r}]}, 
\end{eqnarray}
where
\begin{equation}
h(q_y)= \sum_{{\bf k} \in RBZ} \left[ 2 \cos{k_x}-\cos{k_y}-\cos{(k_y+q_y)} \right]
\end{equation}
%

The resulting charge modulation is shown in Fig. 2 (b).
Notice that the stripe patterns of the charge ordering occur
for ${\bf q}=(0,\pm |q_y|)$ and ${\bf q}=(\pm |q_x|, 0)$.
The amplitude of the charge modulation depends on 
the value of $q_y$ via $h(q_y)$ (for example, $h(\pi/5)= -0.42$) 
and it vanishes at $q_y=0$, which confirms the absence of 
a charge modulation in the d-density wave state.
A similar, but different, relation between the charge density wave
and the spin density wave was discussed in \cite{kivelson} 
and \cite{schulz}. The main difference between the stripe-like 
structure found here and those discussed in the context of 
the spin density wave is that there exists an ``internal''
charge modulation inside each charge stripe in the present case.
%
%

A checker-board pattern of the charge ordering can also occur
if ${\bf q}= (\pm |q_x|,\pm |q_y|)$. The checker-board pattern 
gets distorted as the asymmetry between $q_x$ and $q_y$ increases.

\paragraph{Discussion and Conclusion}

We show, at the mean-field level, that the specific heat jump at the 
transition temperature of the d-density wave order gets weaker when 
the chemical potential is finite. As a result, it becomes more
difficult to determine the existence of the d-density wave state as
the system is away from half-filling. 
The relevance of our results to the cuprates depends on the doping 
dependence of the chemical potential, which is currently not well
understood. There is an indication in La$_{2-x}$Sr$_x$CuO$_4$ that
the chemical potential may be pinned in the underdoped regime.\cite{fuji} 
In this case, our result of the weak specific heat anomaly does not apply.
Our results are more relevant when the chemical 
potential moves sensitively as the doping concentration 
changes, which seems to happen near the optimally 
doped regime.\cite{fuji} Further increase of the chemical potential
leads to a phase transition into the incommensurate 
orbital antiferromagnetic state, where a novel charge ordering occurs.
The charge ordering patterns depend on the nature of the incommensuration
of the current density and can be observed by X-ray scattering. 
Some of the possible charge ordering patterns
have stripe structures similar to, but different from, those discussed 
in the context of the spin density wave in the cuprates.\cite{kivelson,schulz}  

{\it Acknowledgments:}
We are grateful to Sudip Chakravarty and Eugene Demler for helpful discussions.
We thank Kazumi Maki for pointing out the reference \cite{huang92}.  
This work was supported in part by Canadian Institute for 
Advanced Research (H.Y.K. and Y.B.K.) and Alfred P. Sloan Foundation (Y.B.K.).


\begin{references}
\bibitem{iddw} The existence of such a phase due to certain frustration in 
the d-density wave state was also discussed in Refs.\cite{sudip00} 
and \cite{nayak}.
\bibitem{note} The density of state of the tight binding model in two
dimension has a logarithmic singularity at $\epsilon=0$.
However, since we are interested in the mean field results, we have
assumed no anomaly in arbitrary dimension.
\end{references}
\end{document}